\documentclass[aps,pra,superscriptaddress,10pt,twocolumn]{revtex4-2}
\usepackage{amsmath,amssymb,amsfonts}
\usepackage{here}
\usepackage{wrapfig}
\usepackage{comment}
\usepackage{url}
\usepackage{ascmac}
\usepackage{amsthm}
\usepackage{mathrsfs}
\usepackage{siunitx}
\usepackage{hyperref}
\usepackage{mathtools}

\usepackage[utf8]{inputenc}
\usepackage{color}
\usepackage{physics}

\usepackage{braket}
\usepackage{bbold}
\usepackage{graphicx}
\usepackage[normalem]{ulem}
\usepackage{bm}
\hypersetup{colorlinks=true,linkcolor=blue,citecolor=blue,urlcolor=blue}

\begin{document}

\title{Controlling discrete time crystals via single-site operations \\
in zero-field diamond quantum simulators}
\author{Naoya Egawa}
\affiliation{Department of Physics, Tohoku University, Sendai, Miyagi 980-8578, Japan}
\email[egawa.naoya.p4@dc.tohoku.ac.jp]{}

\author{Kaoru Mizuta}
\affiliation{Department of Applied Physics, Graduate School of Engineering, The University of Tokyo, Hongo 7-3-1, Bunkyo, Tokyo 113-8656, Japan}
\affiliation{Photon Science Center, Graduate School of Engineering, The University of Tokyo, Hongo 7-3-1, Bunkyo, Tokyo 113-8656, Japan}
\affiliation{RIKEN Center for Quantum Computing (RQC), Hirosawa 2-1, Wako, Saitama 351-0198, Japan}

\author{Joji Nasu}
\affiliation{Department of Physics, Tohoku University, Sendai, Miyagi 980-8578, Japan}

\date{\today}

\begin{abstract}
Discrete time crystals (DTCs) have emerged as novel nonequilibrium phases of matter that spontaneously break discrete time-translation symmetry in periodically driven systems. Rigorous experimental validation of DTCs, which requires highly controllable quantum simulators, has stimulated extensive research across diverse fields in condensed matter physics and quantum information technologies. Among these advances, DTCs were demonstrated in a hybrid spin register within diamond, comprising a processor spin and surrounding memory spins. However, in conventional strategies involving a bias magnetic field, the field application effectively restricts the controllability of the processor spin. This limitation can be a significant barrier to the next goal of DTCs: achieving multifunctionality through enhanced local controllability. In this study, we theoretically propose multiple DTC protocols through the design of specific single-site control within the entire system. To this end, we consider a concrete model of a diamond-based quantum simulator operating without a bias magnetic field, thereby eliminating the restrictions on the processor spin. Our findings demonstrate that single-site operations enable access to DTCs with multiple distinct features in terms of periodicity and lifetime. Therefore, this approach provides a promising platform for creating diverse DTCs induced by single-site operations.
\end{abstract}

\maketitle

\section{Introduction}
Engineering quantum many-body systems in quantum devices is a central objective in the study of out-of-equilibrium quantum dynamics with potential applications in quantum information technologies.
Thermalization, the irreversible loss of information in isolated quantum systems, has been a major obstacle to such engineering efforts, hindering the classification of nonequilibrium phases of matter~\cite{deutschQuantumStatisticalMechanics1991,srednickiChaosQuantumThermalization1994,rigolThermalizationItsMechanism2008,rigolBreakdownThermalizationFinite2009,moriThermalizationPrethermalizationIsolated2018}. 
To avoid reaching thermal equilibrium, novel nonequilibrium steady states such as many-body localization (MBL)~\cite{oganesyanLocalizationInteractingFermions2007,palManybodyLocalizationPhase2010,serbynLocalConservationLaws2013a,husePhenomenologyFullyManybodylocalized2014,luitzManybodyLocalizationEdge2015,nandkishoreManyBodyLocalizationThermalization2015,abaninColloquiumManybodyLocalization2019,sierantManyBodyLocalizationAge2024} and quantum many-body scars~\cite{bernienProbingManybodyDynamics2017,shiraishiSystematicConstructionCounterexamples2017,turnerWeakErgodicityBreaking2018,moudgalyaThermalizationItsAbsence2019,markUnifiedStructureExact2020,serbynQuantumManybodyScars2021,papicWeakErgodicityBreaking2021,moudgalyaQuantumManyBodyScars2022,chandranQuantumManyBodyScars2023} have been being extensively investigated. 

In periodically driven (Floquet) quantum many-body systems, discrete time crystals (DTCs) are one of the stable nonequilibrium many-body states that spontaneously break discrete time-translation symmetry~\cite{khemaniDefiningTimeCrystals2017,khemaniBriefHistoryTime2019,elseDiscreteTimeCrystals2020,zaletelColloquiumQuantumClassical2023}. The prototypical model was initially proposed by incorporating MBL as one of its elements to prevent the system from heating to infinite temperature~\cite{elseFloquetTimeCrystals2016,khemaniPhaseStructureDriven2016, yaoDiscreteTimeCrystals2017}. Starting from these studies, DTC behavior has been observed across a diverse range of experimental platforms, including trapped ions~\cite{zhangObservationDiscreteTime2017,kyprianidisObservationPrethermalDiscrete2021}, NMR systems~\cite{rovnyObservationDiscreteTimeCrystalSignatures2018,rovny31NMRStudy2018, palTemporalOrderPeriodically2018}, ultracold atoms~\cite{auttiObservationTimeQuasicrystal2018}, superconducting quantum processors~\cite{ippolitiManyBodyPhysicsNISQ2021,freyRealizationDiscreteTime2022,miTimecrystallineEigenstateOrder2022}, and spin ensembles in nitrogen-vacancy (NV) centers~\cite{choiObservationDiscreteTimecrystalline2017, choiProbingQuantumThermalization2019,beatrezCriticalPrethermalDiscrete2023,moonDiscreteTimeCrystal2024}, as well as a single NV center~\cite{randallManybodyLocalizedDiscrete2021}. 
Motivated by these observations of hallmarks of DTCs, highly controllable quantum simulators that provide widely tunable initialization and site-resolved measurement have been of great interest~\cite{ippolitiManyBodyPhysicsNISQ2021}. Our study seeks to advance a paradigm that further exploits controllability, achieving greater functionality through individual manipulation. 
Thus far, previous studies have proposed protocols for realizing the coexistence of a DTC and another phase through the application of regional drives~\cite{sakuraiChimeraTimeCrystallineOrder2021,rahamanTimeCrystalEmbodies2024}. Additionally, other studies have demonstrated that specific single-site operations can induce unique DTC properties~\cite{frantzeskakisTimecrystallineBehaviorCentralspin2023,eulerMetronomeSpinStabilizes2024}. 
As the next goal in DTC research, it is necessary to introduce a comprehensive mechanism for inducing multiple DTCs with distinct properties through single-site manipulation in experimentally feasible systems. 

For this purpose, we focus on the system of a single NV center in diamond, composed of an individually controllable and detectable electron spin and surrounding nuclear spins, which serves as a promising platform for quantum information technologies. The central electron spin offers fast control as a processor qubit~\cite{fuchsGigahertzDynamicsStrongly2009}, while multiple nuclear spins interacting with the electron spin exhibit long coherence times, functioning as memory qubits~\cite{bartlingEntanglementSpinPairQubits2022, bradleyTenQubitSolidStateSpin2019, yangHighfidelityTransferStorage2016}. The hybrid use of these two types of qubits has enabled pioneering and versatile implementations, such as quantum networks~\cite{bernienHeraldedEntanglementSolidstate2013,humphreysDeterministicDeliveryRemote2018,pompiliRealizationMultinodeQuantum2021,rufQuantumNetworksBased2021}, quantum error correction~\cite{waldherrQuantumErrorCorrection2014,taminiauUniversalControlError2014,cramerRepeatedQuantumError2016,undenQuantumMetrologyEnhanced2016,abobeihFaulttolerantOperationLogical2022}, magnetometry~\cite{undenQuantumMetrologyEnhanced2016,zaiserEnhancingQuantumSensing2016,degenQuantumSensing2017,wangRandomizationPulsePhases2019,vorobyovQuantumFourierTransform2021}, and quantum many-body simulators~\cite{randallManybodyLocalizedDiscrete2021}. 
Particularly, a prototypical DTC has recently been successfully realized in the nuclear spins on a single NV center with a bias magnetic field~\cite{randallManybodyLocalizedDiscrete2021, vandestolpeMapping50spinqubitNetwork2024,jungDeepLearningEnhanced2021,abobeihAtomicscaleImaging27nuclearspin2019}.
However, in such a system, the controllability of nuclear spins limits the possible state of the electron spin state during the sequence due to the non-degenerate electron spin level structure by a Zeeman magnetic field~\cite{bradleyTenQubitSolidStateSpin2019,taminiauUniversalControlError2014,kolkowitzSensingDistantNuclear2012,vandersarDecoherenceprotectedQuantumGates2012,taminiauDetectionControlIndividual2012}. 
On the other hand, spin registers operating under a zero magnetic field are also attracting attention for their applications in quantum sensing~\cite{vetterZeroLowFieldSensing2022,lenzMagneticSensingZero2021} and quantum communications~\cite{scharfenbergerAbsorptionbasedQuantumCommunication2015,kosakaEntangledAbsorptionSingle2015,ishidaUniversalHolonomicSingle2018,tsurumotoQuantumTeleportationbasedState2019,sekiguchiGeometricEntanglementPhoton2021,sekiguchiOpticallyAddressableUniversal2022,reyesCompleteBellState2022,kamimakiDeterministicBellState2023}. This situation can remove the constraint on the electron spin, benefiting from the degeneracy of its energy level structure. 
Therefore, a single NV center without a bias magnetic field potentially offers a platform for generating multiple DTCs by fully exploiting the degree of freedom in the electron spin. 

In this paper, we explore protocols that induce multiple DTCs through single-electron spin control in the zero-field diamond quantum simulator. Our protocol fully extends the conventional bias-field strategy~\cite{randallManybodyLocalizedDiscrete2021}, enabling us to alter the overall properties of DTCs by exploiting the electron-spin operational degrees of freedom. 
We also demonstrate that our protocols, which are generated through nontrivial single-site operations, allow DTC behavior to persist for a sufficiently long time against specific perturbations. Moreover, we clarify that this stability arises from the intrinsic characteristics of the NV center. 
Our results shed light on methods for controlling DTCs through the design of single-site operations, particularly emphasizing their realization in the diamond NV center.

This paper is organized as follows. In Sec.~\ref{sec2}, we provide a brief overview of the hybrid spin system. In  Sec.~\ref{sec3}, we describe driving protocols for creating multiple DTCs induced by single-site operations. In Sec.~\ref{sec4}, we present numerical results to analyze the characteristics of the DTCs. In Sec.~\ref{sec5}, we explore analytical factors that characterize the DTCs. Finally, Sec.~\ref{sec6} is devoted to the conclusion and outlook. 

\section{A HYBRID SPIN SYSTEM UNDER ZERO MAGNETIC FIELD}~\label{sec2}
We overview the spin register Hamiltonian operating without a bias magnetic field. 
\begin{figure}
\includegraphics[width=3.4in]{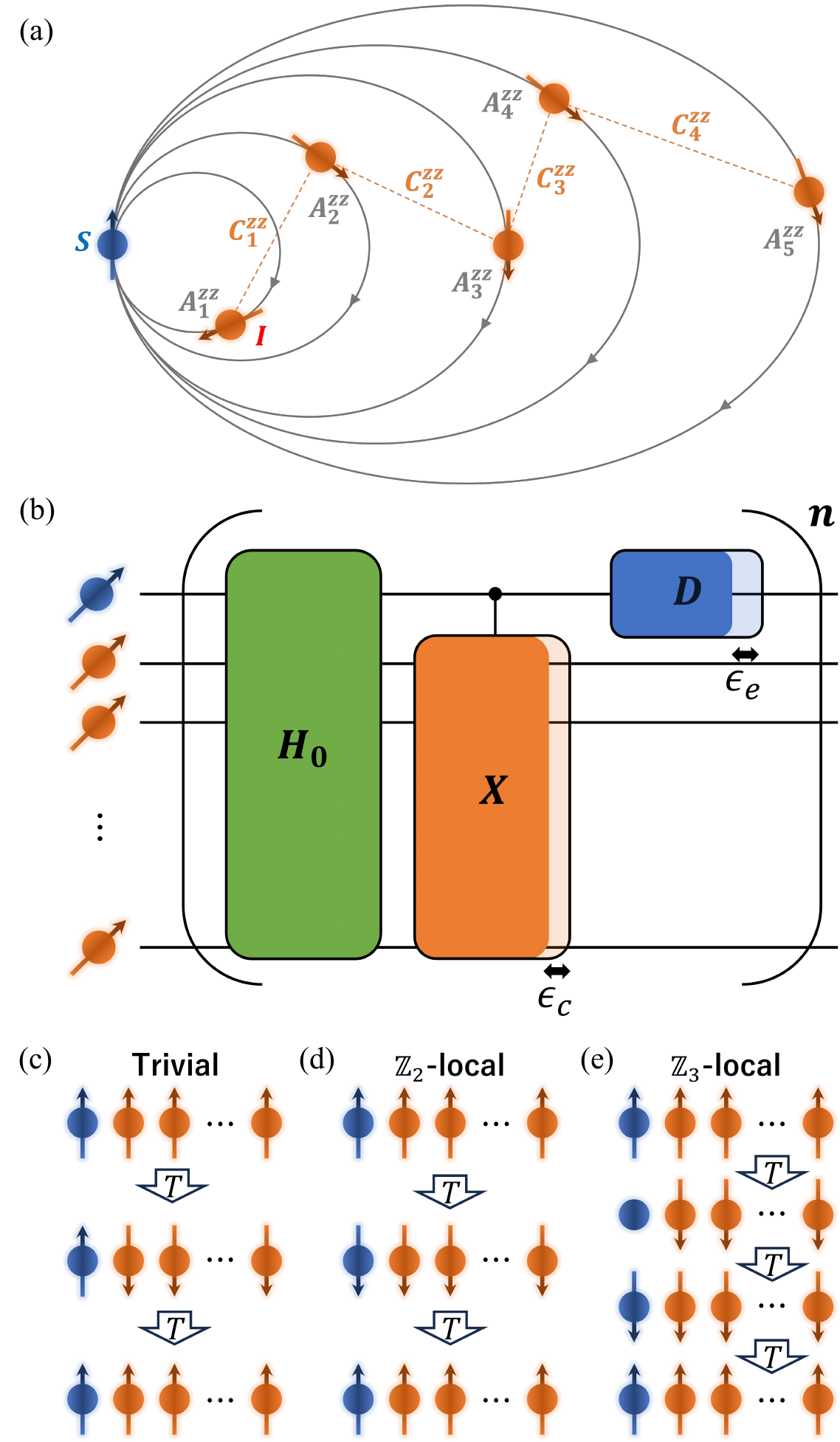}
\caption{(a) Illustration of a hybrid spin system in diamond. A central electron spin (spin-1, blue) is connected to surrounding nuclear spins (spin-1/2, red) through hyperfine interactions represented by $A_{j}^{zz}$ (gray line). Nuclear spins are connected through weaker hyperfine interactions $C_{j}^{zz}$ (dashed orange line). In the absence of a bias magnetic field, the quantization axis of each nuclear spin can be aligned along the orientation of $A_{j}^{zz}$. (b) The gate structure of the Floquet driving protocols consists of three components: a free time evolution (green square), a conditional $X$ gate targeting nuclear spins by controlling the electron spin (orange square), and electron spin driving (blue square). (c-e) Stroboscopic spin dynamics under three different Floquet driving protocols: trivial, $\mathbb{Z}_{2}$-local, and $\mathbb{Z}_{3}$-local. The electron spin quantum number is represented by an up arrow for $+1$, a sphere for $0$, and a down arrow for $-1$. The nuclear spin quantum number is represented by an up arrow for $+1/2$ and a down arrow for $-1/2$.}
\label{dtc_zero_illustration}
\end{figure}
The Hamiltonian describing the electron spin ground state of the negatively charged NV center (NV$^{-}$) and the interacting $^{13}{\rm C}$ nuclear spins is given by~\cite{vandestolpeMapping50spinqubitNetwork2024,wangZerofieldMagnetometryUsing2022}
\begin{align}
    H_{0} = &D_{0}(S^{z})^{2} \nonumber\\
    &+ \sum_j \sum_{\gamma=x,y,z} A_{j}^{\gamma z} S^{\gamma} I_{j}^{z}+ \sum\limits_{j<k} \sum_{\gamma, \gamma^{\prime}=x,y,z} C_{jk}^{\gamma \gamma^{\prime}} I_{j}^{\gamma} I_{k}^{\gamma^{\prime}},
\end{align}
where $S^{x,y,z}$ and $I^{x,y,z}_{j}$ are the spin-$1$ and spin-$1/2$ operators of the electron and nuclear spin, respectively. 
The $^{14}{\rm N}$ nuclear spin terms are assumed to be initialized, and their contributions are ignored~\cite{randallManybodyLocalizedDiscrete2021}. 
$D_{0}$ is the zero-field splitting (ZFS) caused by a crystalline field, 
$A_{j}^{\gamma z} (\ll D_{0})$ is the hyperfine interaction between the electron spin and the $^{13}{\rm C}$ nuclear spin labeled by $j$, and 
$C_{jk}^{\gamma \gamma^{\prime}} \left(\ll D_{0}, A_{j}^{\gamma z}\right)$ is the hyperfine interaction between $^{13}{\rm C}$ nuclear spins $j$ and $k$. 
In this situation, the quantization axis for the electron spin is taken along the orientation of the ZFS, and the quantization axis for each nuclear spin is aligned with the orientation of the individual hyperfine interaction, as shown in Fig.~\ref{dtc_zero_illustration}(a). 
We restrict the interactions between nuclear spins to nearest-neighbor couplings in a one-dimensional chain with open boundary conditions, similar to Ref.~\cite{randallManybodyLocalizedDiscrete2021}. 
In the rotating frame with respect to the ZFS, and applying the secular approximation under the condition $A^{xz}_{j}, A^{yz}_{j} \ll D_{0}$, $C_{j}^{\gamma \gamma^{\prime}} \ll \left\lvert A_{j}^{zz} - A_{j+1}^{zz} \right\rvert (\gamma, \gamma^{\prime}\neq z)$, the Hamiltonian $H_{0}$ is simplified to 
\begin{align}
    H_{0} = \sum\limits_{j} A_{j}^{zz}S^{z}I^{z}_{j} + \sum\limits_{j} C_{j}^{zz} I_{j}^{z} I_{j+1}^{z}. \label{free_evol_hamil}
\end{align}
This hybrid spin system functions as a quantum simulator by applying microwave (MW) or radio-frequency (rf) fields, enabling control of individual spins. 
Due to the frequency selectivity in detecting and controlling nuclear spins, the experimental system size is limited to a few dozen~\cite{abobeihAtomicscaleImaging27nuclearspin2019, bradleyTenQubitSolidStateSpin2019, vandestolpeMapping50spinqubitNetwork2024}. Consequently, similar to the approach taken with the central-spin model~\cite{palTemporalOrderPeriodically2018} and the XXZ connected central-spin model~\cite{frantzeskakisTimecrystallineBehaviorCentralspin2023}, we focus on finite size systems.

\section{Driving protocols for manipulating DTCs}~\label{sec3}
We design three Floquet driving protocols, which we denote as the trivial protocol, the $\mathbb{Z}_{2}$-local protocol, and the $\mathbb{Z}_{3}$-local protocol. The quantum dynamics under these protocols are schematically described by a quantum circuit in Fig.~\ref{dtc_zero_illustration}(b). The first green quantum gate represents the free time evolution of the system given in Eq.~\eqref{free_evol_hamil}. The second gate, shown in orange, represents an rf driving that executes a global nuclear-spin flip conditioned on the electron spin state (see Appendix~\ref{appa}). Note that in this conditional gate, the nuclear spins are manipulated when the electron spin is in the $\{+1, -1\}$ levels; otherwise, the nuclear spins remain unchanged. The final gate, depicted in blue, represents a MW driving that functions as a single-site electron-spin gate. More specifically, the corresponding time-periodic Hamiltonian is written in the following form: 
\begin{align}
    H(t) =
    \begin{cases}
      H_{0} = \sum\limits_{j=1}^{N-1} A_{j}^{zz} S^{z} I^{z}_{j} + \sum\limits_{j=1}^{N-2} C_{j}^{zz} I_{j}^{z} I_{j+1}^{z} \\
       \hfill(0 \leq t \leq \tau), \\
      H_{1} = (\ket{+ 1}\bra{+ 1} + \ket{- 1}\bra{- 1}) \otimes g_{c}(1-\epsilon_{c}) \sum\limits_{j} I_{j}^{x} \\
       \hfill(\tau \leq t \leq T_{1}), \\
      H_{2}(t) = g_{e} (1-\epsilon_{e})(D(t) \otimes \mathbb{1}) \\ \hfill(T_{1} \leq t \leq T).  \label{dtc_protocol}
    \end{cases}
\end{align}
Here, $H_{0}, H_{1}, H_{2}(t)$ are applied for the duration of $\tau$, $t_{1}=T_{1}-\tau$, and $t_{2}=T-T_{1}$, respectively.  The period is $T = \tau + t_{1} + t_{2}$. $N$ denotes the total system size. The operator $\mathbb{1}$ represents the identity on the nuclear spin Hilbert space.  
The rf drive power $g_{c}$ and the MW drive power $g_{e}$ are variables satisfying $g_{c}t_{1}=\pi$ and $g_{e}t_{2}=\pi$. 
$\epsilon_{c}$ and $\epsilon_{e}$ are the systematic $\pi$-pulse rotation errors for the nuclear and electron spins as a perturbation, respectively. 
In the following, we explore the control of DTC behaviors through the engineering of the drive operator $D(t)$ for the single electron spin. 

\subsection{Trivial protocol}
The simplest choice of the drive is 
\begin{align}
      D(t) = 0. 
\end{align}
This represents a trivial operation, functioning as an identity gate on the electron spin within the protocol. The unitary operator describing time evolution for a single Floquet period is given by 
\begin{align}
    U_{\rm tri} = &U_{x, {\rm rf}}(\pi(1 - \epsilon_{c}))U_{A}(\tau)U_{C}(\tau), \label{floquet_tri}
  \end{align}
with
\begin{align}
 U_{A}(\tau) = &\exp(-i \tau \sum\limits_{j} A_{j}^{zz} S^{z} I^{z}_{j} ), \\
 U_{C}(\tau) = &\exp(-i \tau \sum\limits_{j} C_{j}^{zz} I_{j}^{z} I_{j+1}^{z}), \\
 U_{x, {\rm rf}}(\theta) = &(\ket{+1}\bra{+1} + \ket{-1}\bra{-1}) \otimes \exp(-ig_{c}\theta\sum\limits_{j} I_{j}^{x}) \nonumber\\ &+\ket{0}\bra{0}\otimes \mathbb{1}. \label{rf_zero}
\end{align}
By applying the rf driving, we can decouple the nuclear spins from the electron spin while preserving the nuclear-nuclear spin interactions. When the electron spin state is initialized as $\ket{+1}$ or $\ket{-1}$, the period-doubling oscillation of the nuclear spins is ideally observed ($\epsilon_{c}=0$): $\ket{\uparrow} \to \ket{\downarrow} \to \ket{\uparrow}$ [Fig.~\ref{dtc_zero_illustration}(c)]. We note that the electron spin state remains unchanged over time, and one can reduce Eq.~\eqref{dtc_protocol} to the nuclear spin system, which has the same form as the DTC protocol realized in the bias-field diamond spin register. This means that the electron spin acts as a standalone ancilla for nuclear spins. Within a certain range of disorder strength, it is generally recognized that the system exhibits the prototypical MBL-DTC~\cite{randallManybodyLocalizedDiscrete2021,elseFloquetTimeCrystals2016, khemaniPhaseStructureDriven2016, yaoDiscreteTimeCrystals2017}.  

\subsection{\texorpdfstring{$\mathbb{Z}_{2}$}{Z2}-local protocol}
According to Ref.~\cite{bradleyTenQubitSolidStateSpin2019}, in bias-field diamond spin registers, the electron spin must be in the $\ket{-1}$ state to control the nuclear spins using only rf driving. This implies that changing the electron spin state from $m_{s}=-1$ during the dynamics disables the subsequent operation of the nuclear spins. In contrast, without a bias magnetic field, the rf-driven nuclear spin operation functions independently of the electron spin $\{+1, -1\}$ state [see Eq.~\eqref{rf_zero}]. Thus, it becomes possible to incorporate nontrivial electron spin operations into the Floquet driving. This enables the activation of the electron spin as an ancillary system connected to the nuclear spins. To realize another $2T$ periodic dynamics distinct from the previous protocol, we engineer the single-site drive Hamiltonian as (details in Appendix~\ref{appb})
\begin{align}
      D(t) &= \sigma_{x}^{\{+, 0\}}, \\  \sigma_{x}^{\{+, 0\}} &= \ket{+}\bra{0} + \ket{0}\bra{+}, 
\end{align}
where $\ket{+} = (\ket{+1} + \ket{-1})/\sqrt{2}$. This pulse induces a $\pi$-pulse operation on the $\{+1, -1\}$ levels of the electron spin because the $2\pi$ rotation around the $x$-axis in the Bloch sphere spanned by $\{+, 0\}$ corresponds to the $\pi$ rotation around the $x$-axis in the Bloch sphere spanned by $\{+1, -1\}$~\cite{arroyo-camejoRoomTemperatureHighfidelity2014,sekiguchiGeometricSpinEcho2016,sekiguchiDynamicalDecouplingGeometric2019,cerrilloLowFieldNanoNMR2021,vetterZeroLowFieldSensing2022}: 
\begin{align}
U_{x}^{\{+,0\}}(2\pi) &= \exp(-i2\pi\sigma_{x}^{\{+,0\}}/2) \nonumber\\
&= -\ket{+}\bra{+} -\ket{0}\bra{0} + \ket{-}\bra{-} \nonumber\\
&= -\ket{+1}\bra{-1} - \ket{-1}\bra{+1} - \ket{0}\bra{0} \nonumber\\
&= -U_{x}^{\{+1,-1\}}(\pi), 
\end{align}
where $\ket{-} = (\ket{+1} - \ket{-1})/\sqrt{2}$, and 
\begin{align}
U_{x}^{\{+,0\}}(\theta) &= \exp(-i\theta\sigma_{x}^{\{+,0\}}/2), \\
U_{x}^{\{+1,-1\}}(\theta) &= \exp(-i\theta\sigma_{x}^{\{+1,-1\}}/2), \\
\sigma_{x}^{\{+1,-1\}} &= \ket{+1}\bra{-1} + \ket{-1}\bra{+1} + \ket{0}\bra{0}. 
\end{align}
The Floquet unitary is  
\begin{align}
    U_{\rm \mathbb{Z}_{2}} = &U_{x}^{\{+,0\}}(2\pi(1-\epsilon_{e}))U_{x, {\rm rf}}(\pi(1 - \epsilon_{c}))U_{A}(\tau)U_{C}(\tau). \label{floquet_z2}
  \end{align}
When $\epsilon_{c}=\epsilon_{e}=0$, both the electron and nuclear spins oscillate with a period of $2T$: $\ket{+1} \to \ket{-1} \to \ket{+1}$ and $\ket{\uparrow} \to \ket{\downarrow} \to \ket{\uparrow}$ [Fig.~\ref{dtc_zero_illustration}(d)]. 
  
\subsection{\texorpdfstring{$\mathbb{Z}_{3}$}{Z3}-local protocol}
By fully exploiting the spin-1 levels, we can also create the $3T$ periodic behavior. Here, we engineer two consecutive pulses: 
\begin{align}
    D(t) &= 
    \begin{cases}
    \sigma_{x}^{\{-1, 0\}} \quad (T_{1} \leq t \leq T^{\prime}), \\
    \sigma_{x}^{\{+1, 0\}} \quad (T^{\prime} \leq t \leq T), \\
    \end{cases} \\
\sigma_{x}^{\{-1, 0\}} &= \ket{-1}\bra{0} + \ket{0}\bra{-1}, \\
\sigma_{x}^{\{+1, 0\}} &= \ket{+1}\bra{0} + \ket{0}\bra{+1}, 
\end{align}
$T^{\prime} = (T+T_{1})/2$. These two pulses induce $\pi$-pulse operations between the $\{-1, 0\}$ and $\{+1, 0\}$ levels, respectively. The operator drives the $3T$-periodic local unitary operation which is given by~\cite{choiObservationDiscreteTimecrystalline2017,choiProbingQuantumThermalization2019} 
\begin{align}
    &U_{x}^{\{+1, 0\}}(\pi)U_{x}^{\{-1, 0\}}(\pi) \nonumber\\
    &= \ket{+1}\bra{-1} + \ket{-1}\bra{0} + \ket{0}\bra{+1}, 
\end{align}
with 
\begin{align}    
U_{x}^{\{-1, 0\}}(\theta) = \exp(-i\theta\sigma_{x}^{\{-1,0\}}/2), \\
    U_{x}^{\{+1, 0\}}(\theta) = \exp(-i\theta\sigma_{x}^{\{+1,0\}}/2). 
\end{align}
The corresponding Floquet unitary operator is 
\begin{align}
    U_{\rm \mathbb{Z}_{3}} = &U_{x}^{\{+1, 0\}}(\pi(1-\epsilon_{e}))U_{x}^{\{-1, 0\}}(\pi(1-\epsilon_{e})) \nonumber\\
    &\times U_{x, {\rm rf}}(\pi (1- \epsilon_{c})) U_{0}(\tau), \label{floquet_z3}
\end{align}
For $\epsilon_{c}=\epsilon_{e}=0$, the electron spin exhibits a $3T$-periodic oscillation: $\ket{+1} \to \ket{0} \to \ket{-1} \to \ket{+1}$. On the other hand, the nuclear spin oscillation undergoes a different $3T$-period oscillation: $\ket{\uparrow} \to \ket{\downarrow} \to \ket{\downarrow} \to \ket{\uparrow}$ [Fig.~\ref{dtc_zero_illustration}(e)]. The manipulation of the electron spin alters the oscillation behavior of all nuclear spins without changing the direct operations on the nuclear spins. This phenomenon arises because the nuclear spin operation is conditioned on the electron spin state [Eq.~\eqref{rf_zero}]. 

\section{Numerical simulation of real-time dynamics}~\label{sec4}
In this section, we examine the dynamics of the three protocols defined in the previous section and clarify the impact of single-site driving on the time translation symmetry-breaking behavior. We use the QuSpin Python package in the following simulations to compute the time evolution of quantum many-body systems~\cite{weinbergQuSpinPythonPackage2017}. 
In numerical simulations, we assume a sufficiently short MW duration $t_{2}\to 0$ while keeping the MW drive power constant at $g_{e}t_{2}=\pi$~\cite{fuchsGigahertzDynamicsStrongly2009}. According to Ref.~\cite{randallManybodyLocalizedDiscrete2021}, we also assume a similar condition for the rf duration: $t_{1}\to 0$ while keeping the rf drive power as $g_{c}t_{1}=\pi$.

\subsection{Switching DTC behaviors through single-site operation}
\begin{figure*}
\includegraphics[width=7in]{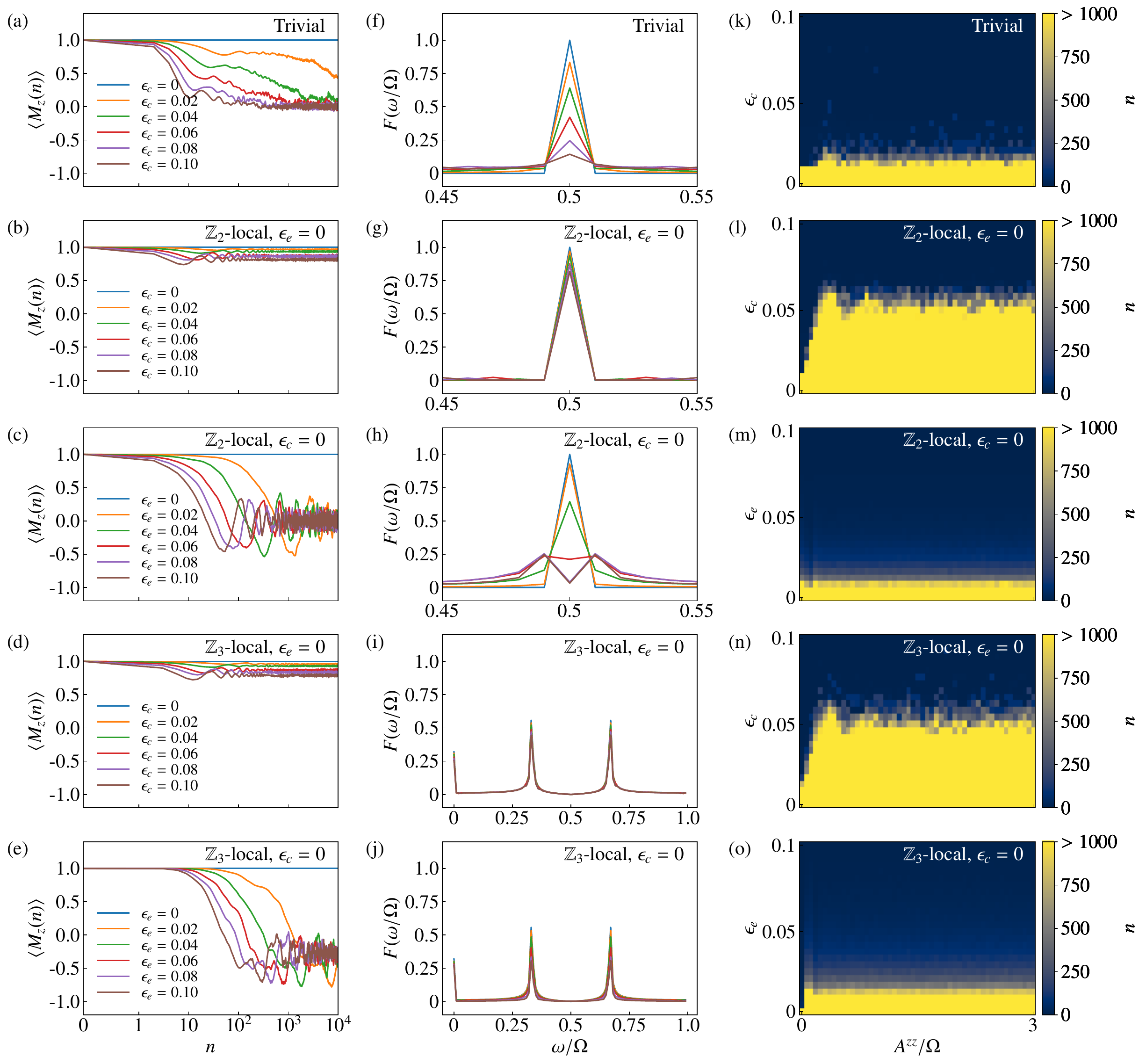}
\caption{Stroboscopic behaviors and lifetime of DTCs in trivial, $\mathbb{Z}_{2}$-local, and $\mathbb{Z}_{3}$-local protocols. We consider a polarized initial state, $\ket{\psi(0)} = \ket{+1} \otimes \ket{\uparrow \uparrow \cdots \uparrow}$, with a system size of $N=6$. Calculations are disorder-averaged over 128 realizations for $A^{zz}_{j} \in [0, 2A^{zz}]$ and $C^{zz}_{j} \in [-C^{zz}, C^{zz}]$, with $C^{zz}/\Omega = 0.12$. (a-e) Stroboscopic dynamics of the nuclear spin magnetization with $A^{zz}/\Omega= 3$. Data points are plotted every two periods for the trivial protocol with (a), for the $\mathbb{Z}_{2}$-local protocol with (b) $\epsilon_{c}$ error, (c) $\epsilon_{e}$ error, and every three periods for the $\mathbb{Z}_{3}$-local protocol with (d) $\epsilon_{c}$ error, (e) $\epsilon_{e}$ error. 
(f-j) Fourier spectra of the stroboscopic dynamics over $100$ Floquet periods ($0 \leq n < 100$) for every Floquet period: (f) Trivial protocol, (g) $\mathbb{Z}_{2}$-local protocol with $\epsilon_{c}$ error, (h) $\mathbb{Z}_{2}$-local protocol with $\epsilon_{e}$ error, (i) $\mathbb{Z}_{3}$-local protocol with $\epsilon_{c}$ errorm, and (j) $\mathbb{Z}_{3}$-local protocol with $\epsilon_{e}$ error. 
(k-o) Signatures of DTC against disorder strength versus rotation error. The color bar represents the maximum revival period $n$ for which the revival probability satisfies $P(nT) = |\braket{\psi(0)|\psi(nT)}|^{2} \geq 0.95$: (k) Trivial protocol, (l) $\mathbb{Z}_{2}$-local protocol with $\epsilon_{c}$ error, (m) $\mathbb{Z}_{2}$-local protocol with $\epsilon_{e}$ error, (n) $\mathbb{Z}_{3}$-local protocol with $\epsilon_{c}$ error, and (o) $\mathbb{Z}_{3}$-local protocol with $\epsilon_{e}$ error.}
\label{dtc_polarized}
\end{figure*}

In the proposed protocols, we have only verified the ideal situations (integrable cases), leaving it uncertain whether time-translation symmetry-breaking oscillations can be realized in the presence of perturbations. To address this ambiguity, we perform numerical calculations to evaluate rigidity, which characterizes the robustness of oscillations against perturbations, and demonstrate that the three proposed protocols indeed exhibit characteristics consistent with DTC behavior.
We start with a polarized initial state $\ket{\psi(0)} = \ket{+1} \otimes \ket{\uparrow \uparrow \cdots \uparrow}$, where
$\ket{\psi(t)}$ denotes the quantum state at time $t$. Subsequently, we compute the stroboscopic dynamics of the nuclear-spin magnetization as follows:
\begin{equation}
    \braket{M_{z}(nT)} = \sum\limits_{j=1}^{N-1} \braket{\psi(0)| (U_{F}^\dagger)^{n} I_{j}^{z} (U_{F})^{n} |\psi(0)}/(N-1),
\end{equation}
where $n$ is the number of Floquet periods. We take $N=6$ as a practically achievable size. 
The hyperfine interactions are randomly sampled as $A_{j}^{zz} \in [0, 2A^{zz}]$ and $C_{j}^{zz}\in [-C^{zz}, C^{zz}]$, where $A^{zz}/\Omega =3$ and $C^{zz}/\Omega = 0.12$. Here, $\Omega=2\pi/T$ is set as a unit of energy. All our results are averaged over $128$ independent realizations of coupling disorder.

To quantitatively evaluate the robustness against rotation errors, 
we compute the time evolution of the nuclear-spin magnetization $\braket{M_{z}(nT)}$, as shown in Figs.~\ref{dtc_polarized}(a)--\ref{dtc_polarized}(e). In these figures, we present the data points only for each return period. In Figs.~\ref{dtc_polarized}(f)--\ref{dtc_polarized}(j), we show a Fourier spectrum $F(\omega)$ of the stroboscopic nuclear-spin magnetization dynamics over a range of periods $n \in [0, 100)$ to illustrate subharmonic oscillation properties. 
For the trivial protocol presented in Fig.~\ref{dtc_polarized}(a), we plot the time evolution of $\braket{M_{z}(nT)}$ for even $n$ under nuclear-spin rotation errors. 
No sign of decay is observed in the absence of nuclear-spin rotation error ($\epsilon_{c}=0$). When an error ($\epsilon_{c}=0.02$) is introduced, the magnetization amplitude decays but exhibits robust behavior against it. As the magnitude of the error increases ($\epsilon_{c}=0.04$--$0.10$), the magnetization approaches zero within the time range shown in the figure. The Fourier spectrum of $\braket{M_{z}(nT)}$ for the trivial protocol [Fig.~\ref{dtc_polarized}(f)] shows that for $\epsilon_c=0$, perfect period-doubling is observed at $\omega=0.5$. This frequency persists even for larger errors, but its peak value gradually decreases. Thus, the numerical analysis of the trivial protocol suggests that DTC behaviors emerge in this protocol.

In contrast, for the $\mathbb{Z}_{2}$-local protocol with nuclear spin rotation errors [Fig.~\ref{dtc_polarized}(b)], we observe a significant enhancement in the lifetime of nuclear spin magnetization dynamics across all values of $\epsilon_{c}$ compared to the trivial protocol. Within the examined timescale, no decay is observed. Furthermore, the subharmonic oscillation frequency [Fig.~\ref{dtc_polarized}(g)] also remains highly stabilized, regardless of $\epsilon_{c}$ values. This stability is achieved solely through the engineering of the manipulation of the electron spin, with no modifications of the rf field applied to the nuclear spin. Thus, the dynamics of the nuclear spin magnetization are effectively altered by intervention in the electron spin. Even in the $\mathbb{Z}_{3}$-local protocol, remarkable stability is observed across a broad range of $\epsilon_{c}$, with no significant decay in amplitude even as error levels increase. We note that this stability persists despite changes in the periodic behavior of the nuclear spins from $2T$ to $3T$ induced by electron spin manipulation [Figs.~\ref{dtc_polarized}(d) and~\ref{dtc_polarized}(i)], underscoring a distinctive characteristic of this protocol. Thus, the numerical results of the nontrivial protocols indicate that atypical DTC behaviors can be realized by the electron spin degrees of freedom in the zero-field diamond quantum simulator. 

We also investigate the robustness of the $\mathbb{Z}_{2}$-local protocol against the rotation error in the electron spin (i.e., $\epsilon_e \neq 0$). We can observe robust period-doubling dynamics at $\epsilon_{e} = 0.02$, although rigidity is weaker compared to the $\epsilon_{c}$ error case. Additionally, for $\epsilon_{c} \geq 0.04$, the magnetization value converges to zero by $n=100$ [Fig.~\ref{dtc_polarized}(c)]. Moreover, a splitting of the oscillation frequency from $\omega = 0.5$ is observed [Fig.~\ref{dtc_polarized}(h)], indicating that the DTC stability against $\epsilon_{e}$ errors is comparatively weaker than that against $\epsilon_{c}$ errors. In the $\mathbb{Z}_{3}$-local protocol under electron spin rotation errors [Figs.~\ref{dtc_polarized}(e) and~\ref{dtc_polarized}(j)], similar stability to that of the $\mathbb{Z}_{2}$-local protocol is observed, with the magnetization value converging to $-1/3$. 

\subsection{Dependence on electron-nuclear spin interaction strength}
The key distinction between the trivial and nontrivial protocols lies in whether the electron spin is manipulated. In other words, the observed change in rigidity due to electron spin manipulation suggests that interactions involving the electron spin influence the rigidity. Thus, by examining the dependence of rigidity on the strength of the electron-nuclear spin interactions, we aim to determine whether the source of stability lies in these interactions. We depict the lifetime of DTC behaviors for the disorder strength of the hyperfine interactions $A_{j}^{zz} \in [0, 2A^{zz}]$ and rotation errors from the polarized initial state $\ket{\psi(0)}$ with $N=6$. The color bar represents the maximum period of $n$ satisfying the condition with the revival probability $P(nT) = |\braket{\psi(0)|\psi(nT)}|^{2} \geq 0.95$ for the return period $n$. This quantity can assess the stability of not only the total nuclear-spin magnetization but also local properties. 
The results shown in Figs.~\ref{dtc_polarized}(k)-\ref{dtc_polarized}(o) indicate that the color map is divided into blue and yellow, suggesting the presence of two regions: one where the DTC behavior is rapidly violated and another where it remains stable over a long period. In the trivial protocol [Fig.~\ref{dtc_polarized}(k)], the stable DTC behavior is observed in a narrow range of $\epsilon_{c}$.

In the $\mathbb{Z}_{2}$-local and $\mathbb{Z}_{3}$-local protocols [Figs.~\ref{dtc_polarized}(l) and~\ref{dtc_polarized}(n)], when the interaction strength $A^{zz}$ is small and comparable to the nuclear spin interaction $C^{zz}$, the upper limit of $\epsilon_{c}$ for rigidity is found to be almost the same as that of the trivial protocol. However, as the interaction strength increases, the DTC region, as indicated in the yellow area, is extended up to $\epsilon_c \simeq 0.05$.  This observation contrasts with the naive expectation that long-range interactions between the electron spin and nuclear spins violate MBL, leading to thermalization~\cite{randallManybodyLocalizedDiscrete2021}. Beyond the boundary between the yellow and blue regions, the system no longer exhibits resilience against $\epsilon_{c}$.

Interestingly, examining the effect of $\epsilon_{e}$ in the nontrivial protocols reveals that the upper bound of the error is lower than that in cases of $\epsilon_{c}$ and is comparable to $\epsilon_{c}$ in the trivial protocol [Figs.~\ref{dtc_polarized}(m) and~\ref{dtc_polarized}(o)]. Moreover, in contrast to the cases with $\epsilon_{c}$, the boundary between the yellow and blue regions appears smeared. 

\subsection{Initial state dependence and disorder effect}
\begin{figure*}
\includegraphics[width=7in]{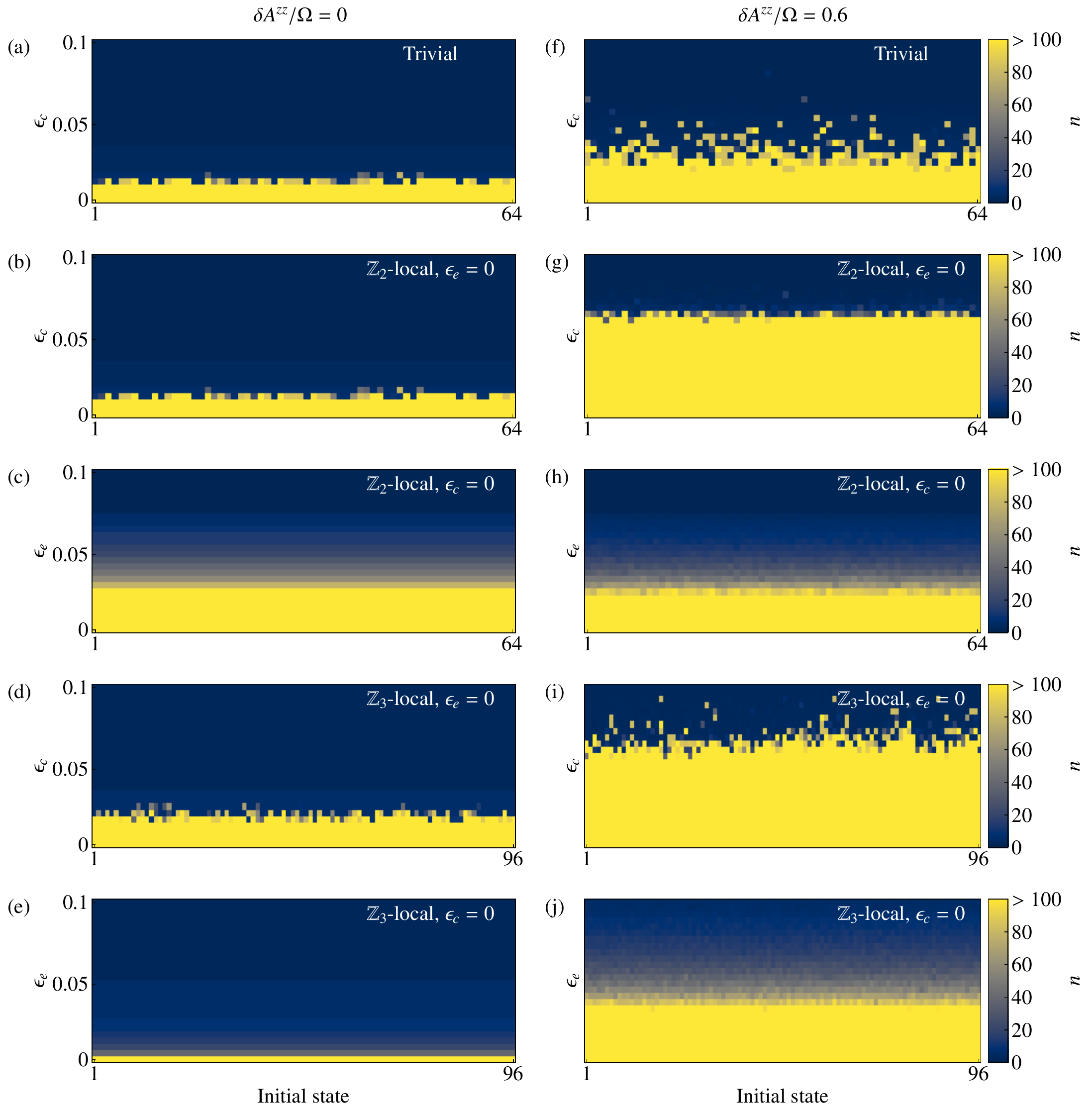}
\caption{Dependence of the period of revival probability threshold on the initial state and disorder. The calculation is averaged over 128 disorder samples for $A^{zz}_{j} \in [A^{zz} - \delta A^{zz}, A^{zz} + \delta A^{zz}]$ and $C_{j}^{zz}\in [-C^{zz}, C^{zz}]$ with $A^{zz}/\Omega = 3$, $C^{zz}/\Omega = 0.12$. The results include the disorder-free case (left column, $\delta A^{zz}/\Omega=0$) and the case with the disorder (right column, $\delta A^{zz}/\Omega=0.6$). We calculate all combinations of the $\{+1, -1\}$ electron spin basis and $\{\uparrow, \downarrow\}$ nuclear spin basis for the (a,f) trivial protocol ($\epsilon_{c}$ error), (b,g) $\mathbb{Z}_{2}$-local protocol ($\epsilon_{c}$ error), and (c,h) $\mathbb{Z}_{2}$-local protocol ($\epsilon_{e}$ error). Additionally, all combinations of the $\{+1, 0, -1\}$ electron spin basis and $\{\uparrow, \downarrow\}$ nuclear spin basis are considered for the (d,i) $\mathbb{Z}_{3}$-local protocol ($\epsilon_{c}$ error) and the (e,j) $\mathbb{Z}_{3}$-local protocol ($\epsilon_{e}$ error). }
\label{initsweep}
\end{figure*}
To distinguish the characteristics of our protocols from the signatures of prethermal DTCs~\cite{ippolitiManyBodyPhysicsNISQ2021,elsePrethermalPhasesMatter2017,mizutaHighfrequencyExpansionFloquet2019,luitzPrethermalizationTemperature2020,machadoLongRangePrethermalPhases2020}, we investigate the time evolution starting from various initial states.
Here, we set the system size to $N=6$ and calculate the maximum revival period $n$ that satisfies $P(nT) \geq 0.95$ for initial product states polarized in the $z$-direction. Note that if the time-evolution exhibits a large $n$, implying that the initial state persists for a long period, its behavior is suggested to differ distinctly from that of prethermal DTCs. More specifically, we analyze all the $2^6$ bases: the combinations of the $\{+1, -1\}$ electron spin basis and $\{\uparrow, \downarrow\}$ nuclear spin basis for the trivial and $\mathbb{Z}_{2}$-local protocols. Additionally, we also consider $3 \times 2^5 = 96$ bases for the initial states of the $\mathbb{Z}_{3}$-local protocol, which are given by the combinations of the $\{+1, 0, -1\}$ electron spin basis and $\{\uparrow, \downarrow\}$ nuclear spin basis. 
Throughout this section, the hyperfine interaction $A^{zz}_{j}$ is randomly chosen from the interval $[A^{zz} - \delta A^{zz}, A^{zz} + \delta A^{zz}]$ where $A^{zz}$ and $\delta A^{zz}$ represent the interaction strength and the amount of disorder, respectively. Similarly, $C_{j}^{zz}$ is randomly chosen from the range $[-C^{zz}, C^{zz}]$, where $C^{zz}$ represents the disorder strength. We choose $A^{zz}/\Omega=3$ and $C^{zz}/\Omega=0.12$. These interactions are averaged over $128$ independent realizations of hyperfine coupling disorder. 

Figure~\ref{initsweep}, we compare the numerical result for the initial state dependence with $\delta A^{zz} /\Omega =0.6$ (disorder), following the setting of this study, and $\delta A^{zz} /\Omega =0$ (disorder-free). The trivial protocol without disorder [Fig.~\ref{initsweep}(a)] shows that rigidity slightly emerges regardless of the initial state. In contrast, in Fig.~\ref{initsweep}(f), we observe enhanced rigidity compared to the disorder-free case for all the initial states in the presence of disorder. This behavior is consistent with the MBL-DTC observed in Ref.~\cite{randallManybodyLocalizedDiscrete2021}. 
In the case without disorder for the $\mathbb{Z}_{2}$-local protocol  [Fig.~\ref{initsweep}(b)], the results exhibit behavior identical to that of the trivial protocol. This indicates that the electron spin operation does not change the DTC properties in the absence of randomness in $A_{j}^{zz}$. Similarly, no significant changes in the DTC properties were observed under the 
$\mathbb{Z}_{3}$-local protocol [Fig.~\ref{initsweep}(d)]. On the other hand, once the disorder is introduced [Figs.~\ref{initsweep}(g) and ~\ref{initsweep}(i)], we find that the rigidity is significantly enhanced compared to the trivial protocol. We note that not only a specific local observable but also the quantum state itself is strongly stabilized against perturbations. 

In the case of electron spin rotation errors, the $\mathbb{Z}_{2}$-local protocol without disordered interactions [Fig.~\ref{initsweep}(c)] demonstrates DTC behavior characterized by a broader change in the maximum revival period. In contrast, the $\mathbb{Z}_{3}$-local protocol [Fig.~\ref{initsweep}(e)] does not exhibit such DTC behavior. We can see that both behaviors are free from fluctuations by $C^{zz}$, and these behavior resemble the mechanism that determines the presence or absence of resonance depending on the value of $A^{zz}/\Omega$ (see Ref.~\cite{frantzeskakisTimecrystallineBehaviorCentralspin2023}). With disordered interactions, the lifetime for the $\mathbb{Z}_{2}$-local protocol remains nearly unchanged compared to the case without disordered interactions [Fig.~\ref{initsweep}(h)]. For the $\mathbb{Z}_{3}$-local protocol with the disorder [Fig.~\ref{initsweep}(j)], a DTC region emerges with a broader change in the maximum revival period similar to Fig.~\ref{initsweep}(f), which is in stark contrast to the absence of such behavior in the disorder-free case. 

\section{Floquet unitary analysis}~\label{sec5} 
In this section, we analyze the Floquet unitary operators and derive the Floquet effective Hamiltonian in our protocols to better understand the key factors contributing to DTCs.

Trivial protocol.---From Eq.~\eqref{floquet_tri}, the square of the Floquet unitary is expressed as
\begin{widetext} 
\begin{align}
    \hat{U}_{\rm tri}^{2} = &\ket{+1}\bra{+1} \otimes U_{x,c}(-\epsilon_{c}\pi)U_{C}(\tau) U_{A,c}^{\dagger}(\tau) U_{x,c}(-\epsilon_{c}\pi) U_{A,c}(\tau) U_{C}(\tau)  \nonumber\\
    &+\ket{-1}\bra{-1} \otimes U_{x,c}(-\epsilon_{c}\pi)U_{C}(\tau) U_{A,c}^{\dagger}(-\tau) U_{x,c}(-\epsilon_{c}\pi) U_{A,c}(-\tau) U_{C}(\tau) \nonumber\\
    &+ \ket{0}\bra{0}\otimes U_{C}(2\tau)     \label{square_tri}, 
  \end{align}
\end{widetext}
where $U_{x,c}(\theta) = \exp(-i \theta \sum\limits_{j}  I_{j}^{x})$ and $U_{A,c}(\theta) = \exp(-i \theta \sum\limits_{j} A_{j}^{zz}I_{j}^{z})$. 
As long as the electron spin state is set to $\ket{+1}$ or $\ket{-1}$, the unitary operator in Eq.~\eqref{square_tri} follows the same structure as the DTC protocol realized in a bias-field NV center~\cite{randallManybodyLocalizedDiscrete2021}, indicating that our protocols fully encompass the case with a bias magnetic field.
Thus, combining the results in Fig.~\ref{initsweep}(f), when the perturbation is small and sufficient randomness is present in $C^{zz}_{j}$ and $A^{zz}_{j}$, we consider dressed spin operators $\tau^{\alpha}_{j} = V I_{j}^{\alpha} V^{\dagger}$ ($\alpha = x,y,z$) with a local unitary $V$~\cite{vonkeyserlingkAbsoluteStabilitySpatiotemporal2016}. The quantity $\tau_{j}^{z}$ serves as the order parameters for the discrete time-translation symmetry breaking: $\{\hat{U}_{\rm tri}, \tau_{j}^{z}\} = 0$, $[\hat{U}_{\rm tri}^{2}, \tau_{j}^{z}] = 0$ for all $j$, and the Floquet unitary exhibits an emergent $\mathbb{Z}_{2}$ symmetry: $[\hat{U}_{\rm tri}, \prod_{j} \tau_{j}^{x}] = 0$~\cite{khemaniDefiningTimeCrystals2017}. 
We note that the component below in Eq.~\eqref{square_tri}, can be calculated as
\begin{align}
&U_{A,c}^{\dagger}(\pm \tau) U_{x,c}(-\epsilon_{c}\pi)U_{A,c}(\pm \tau) \nonumber\\
&=  \exp(i\epsilon_{c}\pi \sum\limits_{j} {\bm n}(\mp A_{j}^{zz}\tau)\cdot {\bm I}_{j}), \label{unitary_hf}
  \end{align}
  where ${\bm n}(\mp A_{j}^{zz}\tau) = \left(\cos(\mp A_{j}^{zz}\tau), \sin(\mp A_{j}^{zz}\tau), 0\right)$.
Therefore, when the nuclear spin rotation error is sufficiently small, i.e., $\epsilon_{c} \ll 1$, and using the Baker-Campbell-Hausdorff formula, the Floquet effective Hamiltonian of Eq.~\eqref{square_tri} can be represented as  
\begin{align}
  &H_{\rm tri}^{2T} \nonumber\\
  &= \sum\limits_{j} \frac{\tau}{T}C_{j}^{zz}I_{j}^{z} I_{j+1}^{z} \nonumber\\
  &- \frac{\epsilon_{c}}{2T}(\ket{+1}\bra{+1} + \ket{-1}\bra{-1}) \otimes \sum\limits_{j}\left(I_{j}^{x} +{\bm n}(\mp A_{j}^{zz}\tau)\cdot {\bm I}_{j}\right),  
  \label{effectivehamil_id}
\end{align}
where the first term corresponds to the stabilizing component, and the second term represents the destabilizing component. For the electron spin state fixed to $\ket{+1}$ or $\ket{-1}$, the effective Hamiltonian is reduced to~\cite{randallManybodyLocalizedDiscrete2021} 
\begin{align}
      &H_{\rm tri}^{2T} \nonumber\\
  &= \sum\limits_{j} \frac{\tau}{T}C_{j}^{zz}I_{j}^{z} I_{j+1}^{z} - \frac{\epsilon_{c}}{2T} \sum\limits_{j}\left(I_{j}^{x} +{\bm n}(\mp A_{j}^{zz}\tau)\cdot {\bm I}_{j}\right). 
\end{align}
This is nothing but the transverse-field Ising model for the nuclear spins since the second term is regarded as a site-dependent field on the plane perpendicular to the $I^{z}$ direction.

$\mathbb{Z}_{2}$-local protocol.---Let us consider the scenario in Eq.~\eqref{floquet_z2}. 
To explore the characteristic stabilization mechanism, we consider the case of the perfect electron spin operation but with a small rotation error for the nuclear spins ($\epsilon_{e}=0, \epsilon_{c} \ll 1$). From Eq.~\eqref{floquet_z2}, the square of the Floquet unitary is expressed as
\begin{widetext}
\begin{align}
    \hat{U}_{\mathbb{Z}_{2},c}^{2} = &\ket{+1}\bra{+1} \otimes U_{x,c}(-\epsilon_{c}\pi)U_{C}(\tau) U_{A,c}(\tau) U_{x,c}(-\epsilon_{c}\pi)U_{A,c}(\tau) U_{C}(\tau) \nonumber\\
    &\quad + \ket{-1}\bra{-1} \otimes U_{x,c}(-\epsilon_{c}\pi)U_{C}(\tau) U_{A,c}(-\tau) U_{x,c}(-\epsilon_{c}\pi)U_{A,c}(-\tau) U_{C}(\tau) \nonumber\\
    &\quad + \ket{0}\bra{0}\otimes U_{C}(2\tau). 
    \label{sr_unitary_z2c}
  \end{align}
\end{widetext}

Compared with Eq.~\eqref{square_tri}, the component corresponding to Eq.~\eqref{unitary_hf} is $U_{A,c}(\pm \tau) U_{x,c}(-\epsilon_{c}\pi)U_{A,c}(\pm \tau)$, meaning that the Hermitian conjugate $\dagger$ for the left $U_{A,c}$ is absent. This originates from the fact that $U_{A}$ is not an operator solely on the nuclear spin; it also operates on the electron spin Hilbert space. Thus, applying the $\pi$-pulse to the electron spin induces the Pauli anti-commutation relation between $U_{A}$ and eliminates the dagger. 
Through high-frequency expansion analysis, the Floquet effective Hamiltonian for Eq.~\eqref{sr_unitary_z2c} is calculated as
  \begin{align}
    H_{\mathbb{Z}_{2},c}^{2T} = &\frac{\tau}{T} \left( \sum\limits_{j} A_{j}^{zz}S^{z} I_{j}^{z} + \sum\limits_{j} C_{j}^{zz}I_{j}^{z} I_{j+1}^{z} \right) \nonumber\\
    &\quad - \frac{\epsilon_{c}\pi}{T}(\ket{+1}\bra{+1} + \ket{-1}\bra{-1})\otimes \sum\limits_{j} I_{j}^{x} \nonumber\\
    &\quad + \frac{1}{T}\mathcal{O}\left(A_{j}^{zz}\tau \epsilon_{c}, C_{j}^{zz}\tau\epsilon_{c}\right), 
    \label{effectivehamil_z2_c}
  \end{align}
where the explicit factor of $A_{j}^{zz}\tau$ appears due to the absence of the dagger. The first term represents a factor for stabilizing the DTC, and the second and third terms correspond to destabilizing factors. We note that as long as $A^{zz}\tau \lesssim 1$, we can neglect the third term.
Specifically, when we prepare the initial electron spin state as $\ket{+1}$ or $\ket{-1}$, the long-range interactions between the electron-nuclear spins, which are naively considered to destroy MBL are replaced by an on-site longitudinal magnetic field applied to the nuclear spins. Consequently, the effective Hamiltonian can be reduced to the nuclear spin system as follows: 
  \begin{align}
    &H_{\mathbb{Z}_{2},c}^{2T} \nonumber\\
    &\simeq \frac{\tau}{T}  \left(\pm  \sum\limits_{j} A_{j}^{zz} I_{j}^{z} + \sum\limits_{j} C_{j}^{zz}I_{j}^{z} I_{j+1}^{z} \right) - \frac{\epsilon_{c}\pi}{T} \sum\limits_{j} I_{j}^{x}, \nonumber\\
  \end{align}
where the sign $\pm$ depends on the choice of the initial electron spin. Thus, the region where $A^{zz} \tau$ takes a small value [left-side region in Fig.~\ref{dtc_polarized}(l)] can be interpreted as indicative of behavior in the MBL regime~\cite{sierantChallengesObservationManybody2022, sierantManyBodyLocalizationAge2024}. Conversely, when $A^{zz}\tau$ takes a sufficiently larger value than $\epsilon_{c}$ in the right-side region of Fig.~\ref{dtc_polarized}(l), this behavior can be regarded as characteristic of the MBL regime with extraordinary stability. This phenomenon cannot be explained by the prototypical DTC framework. 

For $\epsilon_{c}=0, \epsilon_{e} \ll 1$, in this situation, even if we prepare the initial state of the electron spin as the $z$-basis product state, it is impossible to exclude contributions from the electron spin operator, and the errors in the electron spin hinder MBL due to their long-range nature. Unlike the $\epsilon_{c}$ case, the error in the $\pi$-pulse operation within the $\{+,0\}$ space causes leakage out of the $\{+1,-1\}$ space. 
MBL does not persist under this type of perturbation, thereby resulting in a long-range, nonintegrable Floquet effective Hamiltonian as the error increases while gradually disrupting the DTC regime~[Figs.~\ref{dtc_polarized}(m) and \ref{dtc_polarized}(o), Figs.~\ref{initsweep}(h), and~\ref{initsweep}(j)]. 
From these perspectives, we find that local operations on the electron spin influence the symmetry of the system and a change in its rigidity.

$\mathbb{Z}_{3}$-local protocol.--- This protocol can be analyzed in a manner similar to the $\mathbb{Z}_{2}$-local protocol. For $\epsilon_{e}=0, \epsilon_{c} \ll 1$, and $A_{zz}\tau \lesssim 1$, the effective Hamiltonian over a $3T$-period can be derived as follows (details in Appendix~\ref{appc}): 
\begin{align}
&H_{\mathbb{Z}_{3},c}^{3T} \nonumber\\
&\simeq \frac{1}{3T} \left(-2\epsilon_{c}\pi \sum\limits_{j}I_{j}^{x} \pm 2\sum \limits_{j} A_{j}^{zz}\tau I_{j}^{z} + 3\sum\limits_{j} C_{j}^{zz}\tau I_{j}^{z}I_{j+1}^{z} \right), 
\end{align}
where the second term takes a $+$ sign when the initial electron spin state is $\ket{+1}$, and a $-$ sign when the initial electron spin state is $\ket{0}$ or $\ket{-1}$. Thus, the MBL regime again emerges even for the $\mathbb{Z}_{3}$-local protocol and achieves strong rigidity.
The electron spin rotation error during the $\pi$-pulse operation leads to unwanted leakage, resulting in gradual DTC melting. As a result, the population leakage ends up with weaker rigidity compared to nuclear spin rotation errors.

\section{Conclusion}~\label{sec6}
In summary, we have investigated the manipulation of multiple DTCs using single-electron spin control in a zero-field diamond quantum simulator. 
Compared to the conventional bias-field method, we found that the zero-field quantum simulator offers additional degrees of freedom in electron spin levels, which function as an ancilla coupled with nuclear spins. By exploiting these degrees of freedom, single-site operations on the electron spin enable the realization of multiple DTCs with distinct behaviors. Our extended approach builds upon and significantly enhances previous studies on a single NV center, offering a more versatile protocol.
Furthermore, we demonstrated that the DTCs induced by nontrivial single-site operations exhibit significantly enhanced stability due to the characteristic long-range disordered interactions in the single NV center. 
To deepen our understanding of the rich phenomena associated with such unique systems with long-range interactions arising from the presence of a central spin, it is worth conducting further investigations, including analyses of entanglement entropy and eigenvalue spacing distributions~\cite{yaoDiscreteTimeCrystals2017}. Our methodology has potential applications that extend beyond time-crystal phenomena, encompassing a broad range of nonequilibrium dynamics that leverage single-site quantum control capabilities. For instance, our proposed gate structure can be considered as a spin echo~\cite{carrEffectsDiffusionFree1954,souzaRobustDynamicalDecoupling2012} in the context of quantum information technology, our results potentially find applications in sensing through the use of a single NV center~\cite{ieminiFloquetTimeCrystals2024,moonDiscreteTimeCrystal2024}. 

There are several experimental challenges that must be overcome to achieve single-site control of DTCs. For instance, we neglected rotation-axis errors in the electron spin when designing protocols, which arise from environmental $^{13}{\rm C}$ nuclear spins and imperfect initialization of the $^{14}{\rm N}$ nuclear spin. These factors are typically regarded as significant obstacles that could hinder the realization of DTCs. To implement our method, further advancements in dynamical decoupling techniques for the spin-1 V-shaped level structure~\cite{vetterZeroLowFieldSensing2022,zhouRobustHamiltonianEngineering2024,liPhasedGeometricControls2023,violaDynamicalSuppressionDecoherence1998,abobeihOnesecondCoherenceSingle2018,cywinskiHowEnhanceDephasing2008,jiangUniversalDynamicalDecoupling2011,souzaRobustDynamicalDecoupling2012,aliahmedRobustnessDynamicalDecoupling2013} are required. Additionally, developing three-dimensional mapping techniques for weakly coupled nuclear spins in zero-field setup~\cite{randallManybodyLocalizedDiscrete2021, vandestolpeMapping50spinqubitNetwork2024,jungDeepLearningEnhanced2021,abobeihAtomicscaleImaging27nuclearspin2019,taminiauUniversalControlError2014,kolkowitzSensingDistantNuclear2012,vandersarDecoherenceprotectedQuantumGates2012} is essential from both theoretical and experimental perspectives.

\section{Acknowledgement}
The authors thank H. Kosaka, A. Ono, and A. Sakurai for fruitful discussions. Parts of the numerical calculations were performed in the supercomputing
systems in ISSP, the University of Tokyo. This work was partially supported by the WISE Program for AI Electronics in Tohoku University. K. M. was supported by JST PRESTO Grant No.~JPMJPR235A and JSPS KAKENHI Grant No.~JP24K16974. J. N. was supported by JSPS KAKENHI Grant No.~JP20H00122, JP22H01175, JP23H01129, JP23H04865, JP24K00563. 

\begin{widetext} 
\appendix
\section{Radio-frequency driving for nuclear spins}~\label{appa}
In this appendix, we derive the global nuclear spin operation using multi-frequency rf driving, following Ref.~\cite{randallManybodyLocalizedDiscrete2021}. The Hamiltonian with an rf-field is described as
    \begin{align}
    H(t) = \sum\limits_{j} A_{j}^{zz} S^{z} I^{z}_{j} + \sum\limits_{j} C_{j}^{zz} I_{j}^{z} I_{j+1}^{z} + H_{\rm rf}(t), 
\end{align}
with 
\begin{align}
        H_{\textrm{rf}}(t) &= \sum_{j} \Omega(t) I_{j}^{x}, \\
        \Omega(t) &= \sin^{2}\left( \frac{\pi t}{t_{p}} \right) \sum_{k} \Omega_{k} \cos \left( \omega_{k} t + \phi_{k} \right).  
\end{align}
We have omitted the off-diagonal nuclear-nuclear spin interaction terms by applying the secular approximation, assuming that $C_{j}^{x,y} \ll A_{j}^{zz}$. The sine-squared pulse envelope is used for the multi-frequency pulses to reduce crosstalk between nuclear spins~\cite{randallManybodyLocalizedDiscrete2021}. 

In the rotating frame of $H_{R} = \sum\limits_{j} A_{j}^{zz} S^{z} I^{z}_{j}$, the Hamiltonian is written as
\begin{align}
    H^{\prime}(t)
    = & \sum\limits_{j} C_{j}^{zz} I_{j}^{z} I_{j+1}^{z} + \sum\limits_{j}\sum\limits_{k} \Omega_{k} \sin^{2}\left( \frac{\pi t}{t_{p}} \right) \left[\left(\frac{e^{i(\omega_{k} t + \phi_{k} )} + e^{-i(\omega_{k} t + \phi_{k} )}  }{2}  \right) e^{iH_{R}t}I_{j}^{x}e^{-iH_{R}t}\right], 
\end{align}
The hyperfine interaction $C_{j}^{zz}$ is disregarded under the condition $C_{j}^{zz} \ll \Omega_{j}$. 
By tuning the rf frequency as $\omega_{k}$ to $A_{k}^{zz}$, the Hamiltonian is represented as
\begin{align}
    H^{\prime}(t) &= \sum\limits_{k} \Omega_{k} \sin^{2} \left(\frac{\pi t}{t_{p}}\right) \left[(\ket{+1}\bra{+1} + \ket{-1}\bra{-1}) \otimes {\bm n}\left(\pm \phi_{k} \right) \cdot {\bm I} + \ket{0}\bra{0} \otimes  \sum_{j} \cos \left( \omega_{k} t + \phi_{k} \right) I_{j}^{x}\right]  \nonumber \\
    &= \sum\limits_{k} \Omega_{k} \sin^{2} \left(\frac{\pi t}{t_{p}}\right) \left[(\ket{+1}\bra{+1} + \ket{-1}\bra{-1}) \otimes {\bm n}\left(\pm \phi_{k} \right) \cdot {\bm I} + \ket{0}\bra{0} \otimes  \mathbb{1}\right]. 
\end{align}
The last line of the above equation implies that when the electron spin state is $\ket{0}$, the identity operator is applied to the nuclear spin, which is due to the absence of energy gaps to tune the radio-frequency. 
The form of the unitary operator is then described as 
\begin{align}
    U^{\prime}(t_{p}) = &\exp\left( -i  \int_{0}^{t_{1}} H^{\prime}(t)  \,dt \right) \nonumber\\
= &(\ket{+1}\bra{+1} +\ket{-1}\bra{-1}) \otimes \exp\left( -i \frac{t_{1}}{2}\sum\limits_{k}\Omega_{k} {\bm n}\left(\pm \phi_{k} \right) \cdot {\bm I} \right) + \ket{0}\bra{0} \otimes \mathbb{1}. 
\end{align}
Going back to the lab frame, we obtain  
\begin{align}
    U(t_{1}) &= e^{-i \sum\limits_{j} A_{j}^{zz} S^{z} I^{z}_{j}t_{1} }U^{\prime}(t_{1}) \nonumber\\
    &= \ket{+1}\bra{+1} \otimes \exp\left( -i \frac{t_{1}}{2}\sum\limits_{j} A_{j}^{zz}I_{j}^{z} \right) \exp\left( -i \frac{t_{1}}{2}\sum\limits_{k} \frac{\Omega_{k}}{2} {\bm n}\left(\phi_{k}\right) \cdot {\bm I} \right) \nonumber\\
    &\quad + \ket{-1}\bra{-1} \otimes \exp\left( +i \frac{t_{1}}{2}\sum\limits_{j} A_{j}^{zz}I_{j}^{z} \right) \exp\left( -i \frac{t_{1}}{2}\sum\limits_{k} \frac{\Omega_{k}}{2} {\bm n}\left(-\phi_{k}\right) \cdot {\bm I} \right) \\
    &\quad + \ket{0}\bra{0} \otimes \mathbb{1}. 
\end{align}
By defining the rotation angle as $\theta_{k} \coloneqq t_{1}\Omega_{k} /2$, 
assuming short rf-pulse condition\ ($t_{1} \to 0$), and setting the initial phases $\phi_{k} = 0$ and $\theta_{k} = \theta$, the unitary operator is finally described as 
\begin{align}
    U(t_{1}) = &(\ket{+1}\bra{+1} + \ket{-1}\bra{-1} ) \otimes U_{x,c}(\theta) + \ket{0}\bra{0}\otimes \mathbb{1}, 
\end{align}
which corresponds to Eq.~\eqref{rf_zero} in the main text. This indicates that nuclear spins are simultaneously rotated around the $x$-axis when the electron spin is in the $\{+1, -1\}$ levels; otherwise, the nuclear spins are unchanged. 
This results in a conditional operation for nuclear spins dependent on the electron spin state. 
Such behavior suggests the potential to expand the functionality of the NV center as a hybrid spin register.

\section{Microwave driving for electron spin}~\label{appb}
In this appendix, we derive the electron spin $\pi$-pulse operator using microwave driving. The electron spin Hamiltonian operating under a MW field is described as 
\begin{align}
    H = D_{0}S_{z}^{2} + H_{\rm mw}(t),
\end{align}
with
\begin{align}
H_{\rm mw}(t) = \Omega_{x}\cos (\omega_{\rm mw}t + \phi)S_{x}, 
\end{align}
where $\Omega_{x}, \omega_{\rm mw}$, and $\phi$ are the amplitude, frequency, and initial phase of the MW driving. 
In the rotating frame of the ZFS, by tuning the MW frequency as $\omega_{\rm mw}$ to $D_{0}$, we obtain the following Hamiltonian:  
    \begin{align}
    H^{\prime} \approx \frac{\Omega_{x}}{2}{\bm n}\left(\phi \right) \cdot {\bm \sigma}^{\{+,0\}}, 
\end{align}
where ${\bm n}(\phi) = (\cos\phi, \sin\phi, 0)$. 
The $2\pi$ rotation for $\phi=0$ corresponds to
the $\pi$ rotation around the $x$-axis in the Bloch sphere spanned by $\{+1, -1\}$, written in the main text. The $\pi$ rotation in the $\{+1, 0\}$ and $\{-1, 0\}$ levels is also achieved through the application of a polarized MW-field~\cite{nagataUniversalHolonomicQuantum2018}. In practical experimental situations, the surrounding nuclear spins coupled with the electron spin can be a major source of rotation angle errors. Therefore, robust decoupling pulses need to be combined~\cite{vetterZeroLowFieldSensing2022,zhouRobustHamiltonianEngineering2024,liPhasedGeometricControls2023,violaDynamicalSuppressionDecoherence1998,abobeihOnesecondCoherenceSingle2018,cywinskiHowEnhanceDephasing2008,jiangUniversalDynamicalDecoupling2011,souzaRobustDynamicalDecoupling2012,aliahmedRobustnessDynamicalDecoupling2013}.

\section{Analysis of \texorpdfstring{$\mathbb{Z}_{3}$}{Z3}-local Stroboscopic dynamics}~\label{appc}
In this section, we show that the rigidity of the $\mathbb{Z}_{3}$-local stroboscopic dynamics has a similar structure to that of the $\mathbb{Z}_{2}$-local protocol. From Eq.~\eqref{floquet_z3}, the $3T$ Floquet unitary of the $\mathbb{Z}_{3}$-local stroboscopic dynamics in the presence of nuclear-spin rotation errors is given by
\begin{align}
    (U_{\mathbb{Z}_{3},c})^{3} = &\ket{+1}\bra{+1} \otimes U_{x,c}(\pi(1 - \epsilon_{c})) U_{A,c}(-\tau)U_{C}(\tau) U_{C}(\tau) U_{x}(\pi(1 - \epsilon_{c}))U_{A,c}(\tau)U_{C}(\tau) \nonumber\\
    &+ \ket{0}\bra{0} \otimes U_{x,c}(\pi(1 - \epsilon_{c})) U_{A,c}(\tau) U_{C}(\tau) U_{x,c}(\pi(1 - \epsilon_{c}))U_{A,c}(-\tau)U_{C}(\tau)U_{C}(\tau) \nonumber\\
    &+ \ket{-1}\bra{-1} \otimes U_{C}(\tau)U_{x,c}(\pi(1 - \epsilon_{c})) U_{A,c}(\tau)U_{C}(\tau)U_{x,c}(\pi(1 - \epsilon_{c}))U_{A,c}(-\tau)U_{C}(\tau). 
\end{align}
The three terms in $(U_{\mathbb{Z}_{3},c})^{3}$ are summarized as follows: 
\begin{align}
    \textrm{1st term} &= \ket{+1}\bra{+1} \otimes U_{x,c}(-\epsilon_{c}\pi)U_{x,c}(\pi) U_{A,c}(-\tau)U_{x,c}(\pi)U_{C}(\tau) U_{C}(\tau) U_{x,c}(-\epsilon_{c}\pi)U_{C}(\tau) U_{A,c}(\tau) \nonumber\\
    &= \ket{+1}\bra{+1} \otimes U_{x,c}(-\epsilon_{c}\pi)  U_{C}(2\tau) U_{A,c}(\tau) U_{x,c}(-\epsilon_{c}\pi) U_{A,c}(\tau)U_{C}(\tau), 
\end{align}
\begin{align}
    \textrm{2nd term} = \ket{0}\bra{0} \otimes U_{x,c}(-\epsilon_{c}\pi) U_{C}(\tau)U_{A,c}(-\tau)U_{x,c}(-\epsilon_{c}\pi)U_{A,c}(-\tau) U_{C}(2\tau),  
\end{align}
and 
\begin{align}
    \textrm{3rd term} = \ket{-1}\bra{-1} \otimes U_{x,c}(-\epsilon_{c}\pi) U_{C}(2\tau)U_{A,c}(-\tau)U_{x,c}(-\epsilon_{c}\pi)U_{A,c}(-\tau) U_{C}(\tau).   
\end{align}
These terms indicate that the unitary operator on the nuclear spin Hilbert space exhibits a structure analogous to Eq.~\eqref{sr_unitary_z2c} in the main text. Hence, the hyperfine interaction $A_{j}^{zz}$ explicitly appear in the effective Hamiltonian of $H_{\mathbb{Z}_{3}, c}^{3T}$. More specifically, the form of the effective Hamiltonian is given by the high-frequency expansion: 
\begin{align}
H_{\mathbb{Z}_{3}, c}^{3T} = &\ket{+1}\bra{+1} \otimes \frac{1}{3T} \left(-2\epsilon_{c}\pi \sum\limits_{j}I_{j}^{x} + 2\sum \limits_{j} A_{j}^{zz}I_{j}^{z}\tau + 3\sum\limits_{j} C_{j}^{zz}I_{j}^{z}I_{j+1}^{z}\tau \right) \nonumber\\
&+ \ket{0}\bra{0} \otimes \frac{1}{3T} \left(-2\epsilon_{c}\pi \sum\limits_{j}I_{j}^{x} - 2\sum \limits_{j} A_{j}^{zz}I_{j}^{z}\tau + 3\sum\limits_{j} C_{j}^{zz}I_{j}^{z}I_{j+1}^{z}\tau\right) \nonumber\\
&+ \ket{-1}\bra{-1} \otimes\frac{1}{3T} \left(-2\epsilon_{c}\pi \sum\limits_{j}I_{j}^{x} - 2\sum \limits_{j} A_{j}^{zz}I_{j}^{z}\tau + 3\sum\limits_{j} C_{j}^{zz}I_{j}^{z}I_{j+1}^{z}\tau \right) \nonumber\\
&+ \mathcal{O}(A^{zz}_{j}\epsilon_{c}\tau, C^{zz}_{j}\epsilon_{c}\tau). 
\end{align}
For the case with electron-spin rotation errors, the Hamiltonian $H_{\mathbb{Z}_{3}, e}^{3T}$ corresponds to the ergodic effective Hamiltonian, leading to a gradual breakdown of DTC.

\end{widetext}

\bibliography{dtc_nv}

\end{document}